\documentclass[12pt]{article}
\usepackage{deluxetable}
\usepackage{scicite}
\usepackage{times}
\usepackage{psfig}
\usepackage{abbrevs}
\usepackage[bf]{caption2}

\newenvironment{sciabstract}{%
\begin{quote} \bf}
{\end{quote}}

\newcounter{lastnote}

\newcommand{\pbdot}{\ensuremath{\dot{P}_{\mathrm b}}}
\newcommand{\pb}{\ensuremath{P_{\mathrm b}}}

\newcommand{\kms}{\ensuremath{\mathrm{km}\ \mathrm{s}^{-1}}}
\newcommand{\dpsr}{PSR J0737--3039A/B}

\title{Implications of a VLBI Distance to the Double Pulsar J0737--3039A/B}

\author{
A.T. Deller$^{1\ast}$, M. Bailes$^{1}$, S.J. Tingay$^{2}$\\
\normalsize{$^{1}$Centre for Astrophysics and Supercomputing}\\
\normalsize{Swinburne University of Technology}\\
\normalsize{Mail H39, P.O. Box 218, Hawthorn, VIC 3122, Australia}\\
\normalsize{$^{2}$Curtin Institute of Radio Astronomy}\\
\normalsize{Curtin University of Technology, Bentley, WA, Australia}\\
\\
\normalsize{$^\ast$To whom correspondence should be addressed. E-mail:  adeller@astro.swin.edu.au.}
}
\date{}
\begin{document}
\baselineskip24pt

\maketitle

\begin{sciabstract}
The double pulsar J0737--3039A/B is a unique system with which to test
gravitational theories in the strong--field regime.  However, the accuracy of such tests 
will be limited by knowledge of the distance 
and relative motion of the system.  Here we present very long baseline interferometry 
observations which reveal that the distance to PSR J0737--3039A/B is
$1150^{+220}_{-160}$\ pc, more than double previous estimates, and
confirm its low transverse velocity ($\sim$9 km s$^{-1}$). 
Combined with a decade of pulsar timing, these results will allow tests of gravitational radiation 
emission theories at the 0.01\% level, putting stringent constraints on
theories which predict dipolar gravitational radiation.  They also allow insight
into the system's formation and the source of its high--energy emission.
\end{sciabstract}

The double pulsar system J0737--3039A/B \cite{burgay03a,lyne04a}
is one of eight known double neutron star (DNS) systems, and the only system in which both 
neutron stars are visible as pulsars.  The `A' pulsar, which has been spun--up by accretion 
(``recycled") to a period $P=22.7$\,ms, was discovered first; the non--recycled 
`B' pulsar ($P=2.77$\,s) was found during follow--up timing observations.
\dpsr\ is the most relativistic known DNS system.
It has an orbital period of 2.5 hours and a coalescence time (due to orbital energy loss to 
gravitational radiation) of 85 Myr.  Compared to most DNS systems, it has
a low eccentricity (0.08) and is thought to possess a low transverse velocity 
[10 km s$^{-1}$; \cite{kramer06a}], which is difficult to explain in standard models
of pulsar formation.  Pulsar timing of \dpsr\ produces results consistent with the
theory of general relativity (GR) at the 0.05\% level \cite{kramer06a}. 

The distance to \dpsr\ is estimated using the pulsar dispersion measure
(DM) and models of the ionized component of the Galaxy \cite{taylor93a,cordes02a}.
However, distances derived in this manner have been shown to be in error by a factor
of two or more for individual systems [e.g. PSR B0656+14; \cite{brisken03a}].
A more accurate distance to the \dpsr\ system is needed in order to determine the contribution of 
kinematic effects to pulsar timing, whose uncertainty would otherwise limit the precision
GR tests could achieve; to establish the source of the 
x--ray emission from the system \cite{mclaughlin04a,possenti08a}; and to refine the
estimated radio luminosity.

\begin{figure}
\begin{center}
\begin{tabular}{c}
\psfig{file=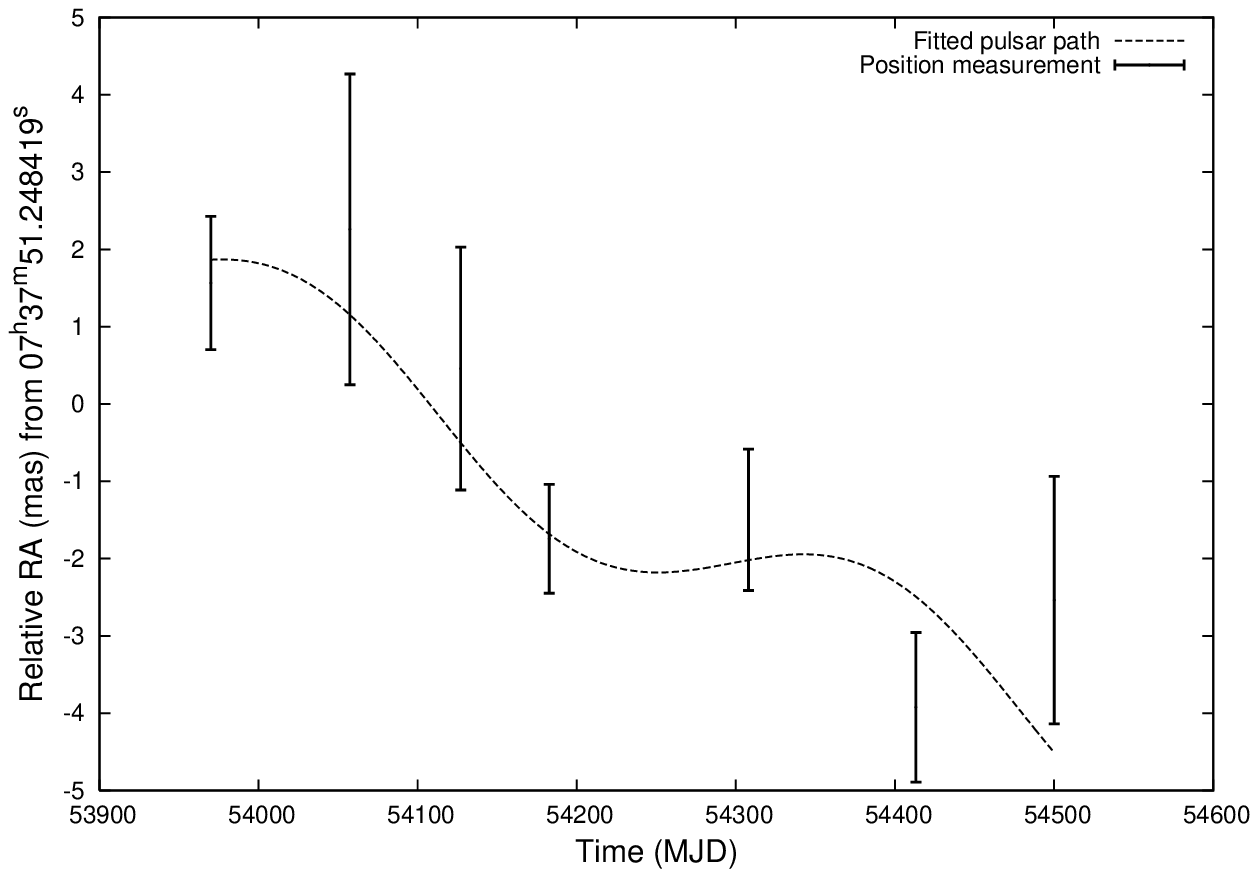, clip=} \\
\psfig{file=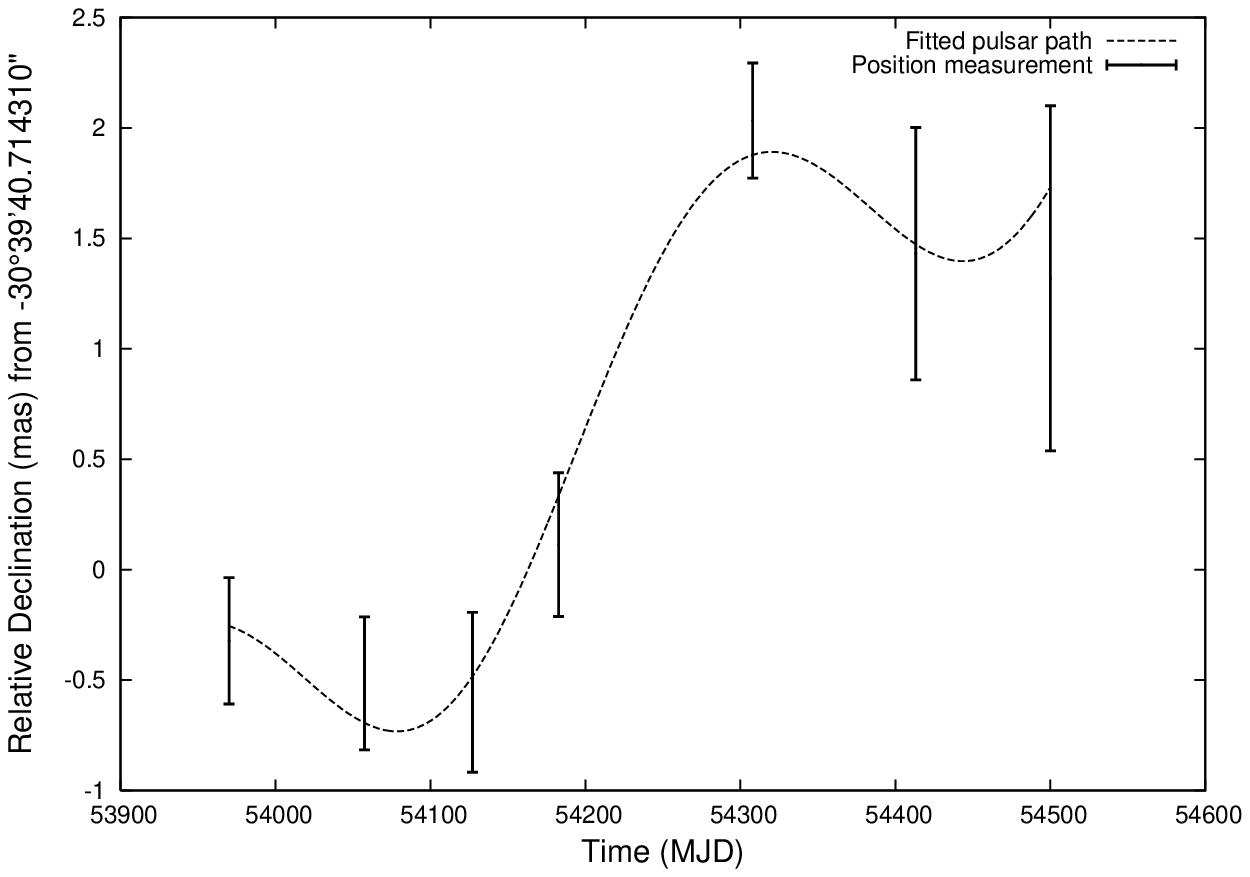, clip=} \\
\end{tabular}
\caption{Motion of \dpsr\ plotted against time. {\bf (Top)}  Offset in right ascension; {\bf (Bottom)}
Offset in declination.  The best fit (dashed line) is 
overlaid on the measured positions.  The reduced chi--squared of the fit was 0.79, implying that 
the measurement errors (and thus the errors on fitted parameters) might be overestimated and
hence conservative.}
\label{fig:0737fit}
\end{center}
\end{figure}

We used the Australian Long Baseline Array (LBA) to obtain a model--independent 
distance to \dpsr, through the direct measurement of its annual geometric parallax.  
We have made seven very long baseline interferometry (VLBI) observations of
\dpsr\ over an 18 month period between August 2006 and February 2008, which allowed us to 
fit the system's position, proper motion and parallax [for more details see \cite{scienceonline}, 
Table~\ref{tab:vlbiresults} and Fig.~\ref{fig:0737fit}; all errors are 1$\sigma$\ unless otherwise
stated].

The distance to \dpsr\ is $1150^{+220}_{-160}$~pc, which is inconsistent with previous DM--based 
estimates of 570~pc \cite{taylor93a} and 480~pc\cite{cordes02a}. Our measurement is
more significant than the distance of $333^{+667}_{-133}$~pc obtained through a marginal timing 
parallax detection \cite{kramer06a}, which differed by approximately 1$\sigma$\ from our result.  
The revised value of electron density along the line of sight to \dpsr\ is 0.043 cm$^{-3}$,
suggesting that the influence of the Gum nebula (a large, ionized hydrogen region between the
Earth and \dpsr) along this line of sight is less than originally thought.
The proper motion of \dpsr\ is $4.37\pm0.65$\ mas yr$^{-1}$, consistent with a previous timing 
measurement of $4.2 \pm 0.6$\ mas yr$^{-1}$ \cite{kramer06a}.

We used the revised distance to \dpsr\ to estimate the precision attainable for testing
predictions of gravitational wave emission from the system.
In order to compare the rate of change of orbital period (\pbdot) due
to the loss of energy to gravitational radiation with the GR prediction, the observed \pbdot\ must
be measured as accurately as possible [the current significance is 70\,$\sigma$ \cite{kramer06a}], 
and contributing factors to \pbdot\ other than GR must be estimated and subtracted as accurately as 
possible. For \dpsr, the major contributing factors are differential Galactic rotation 
($\pbdot^{\mathrm{rot}}$), acceleration towards the plane of the Galaxy ($\pbdot^{\mathrm{z}}$), 
and the apparent acceleration caused by the transverse motion of the system 
[the Shklovskii effect $\pbdot^{\mathrm{Shk}}$; 
\cite{shklovskii70a}], to which we will refer collectively as galactic and kinematic contributions
($\pbdot^{\mathrm{gk}}$).  For the newly-calculated distance of 1150 pc, the magnitude of 
these effects is calculated using Equations~2.12 and 2.28 from \cite{damour91a} as:

\begin{eqnarray}
\frac{\pbdot^{\mathrm{rot}}}{\pb} & = & -\frac{v_{0}^{2}}{c\,R_{0}}\times\left( \cos{l} + \frac{\beta}{\sin^{2}{l}+ \beta^{2}} \right) \label{eq:Pbdotrot}\\
\frac{\pbdot^{\mathrm{z}}}{\pb} & = & -\frac{K_{z}}{c}\ \sin{b}\\
\frac{\pbdot^{\mathrm{Shk}}}{\pb} & = &  \frac{\mu^{2}d}{c} 
\label{eq:Pbdotshk}
\end{eqnarray}

\noindent where $l$, $b$\ and $z$\ are galactic longitude, latitude and height respectively,
$d$\ and $\mu$ are pulsar distance and proper motion, 
$R_{0}$\ and $v_{0}$\ are the Galactic radius and speed of the 
Solar System [taken to be 7.5 kpc and 195 \kms\ respectively \cite{dias05a}], 
$K_{z}$\ is the vertical gravitational potential of the Galaxy [taken from 
\cite{holmberg04a} as 0.45 km$^{2}$\,s$^{-2}$\,pc$^{-1}$ at the height of \dpsr] 
and $c$\ is the speed of light.  The dominant uncertainties are  
in $d$\ (16\%), $\mu$\ (15\%), and $R_{0}$, $v_{0}$\ and $K_{z}$, whose errors are 
estimated at $\sim$10\%.

Equations~(\ref{eq:Pbdotrot})-(\ref{eq:Pbdotshk}) give 
$\pbdot^{\mathrm{rot}}/\pb = (-4.3\pm 0.7) \times10^{-20}$\,s$^{-1}$, 
$\pbdot^{\mathrm z}/\pb = (3.8 \pm 0.8)\times10^{-21}$\,s$^{-1}$, and 
$\pbdot^{\mathrm{Shk}}/\pb = (5.3 \pm 1.8)\times10^{-20}$\,s$^{-1}$.  Combining these terms gives
$\pbdot^{\mathrm{gk}}/\pb = (1.5 \pm 2.1) \times 10^{-20}$\,s$^{-1}$, and multiplying by the observed
orbital period $\pb = 8834.5\,$s \cite{kramer06a} yields the net effect of these terms on
the observed orbital period derivative: $\pbdot^{\mathrm{gk}} = (1.3\pm1.8)\times10^{-16}$.

These contributions to the orbital period derivative are four orders of magnitude below the 
GR contribution, and two orders of magnitude below the current measurement error
[$\pbdot^{\mathrm{obs}} = (-1.252\pm0.017)\times10^{-12}$; \cite{kramer06a}].
Thus, with the current accuracy in the measurement of distance and transverse velocity,
GR tests can be made to the 0.01\% level with \dpsr\ using \pbdot.
However, about ten years of further precision timing will be 
required to reach this point.  Measuring \pbdot\ at
this level will place stringent requirements on the class of gravitational theories which predict
significant amounts of dipolar gravitational radiation, exceeding the best Solar System tests 
\cite{kramer06a}. Measuring the moment of inertia of pulsar A, however, would require 
another order of magnitude improvement in the measurement precision of \pbdot\ \cite{kramer06a}.
In the near future, additional VLBI and/or timing measurements can be expected to reduce the error
in both the distance and velocity of \dpsr\ below 10\%; however, even with negligible
error in these parameters, the existing accuracy of measurements of $R_{0}$, $v_{0}$\ and $K_{z}$
would limit the accuracy of \pbdot\ measurements in this system to 0.004\%.  To attain the $10^{-5}$
precision necessary to measure the neutron star moment of inertia, 
the constants $R_{0}$ and $v_{0}$ must be measured to a precision approaching 1\%.

X--ray observations of \dpsr\ show that most of the x--ray emission from the system is modulated 
at the spin period of the A pulsar \cite{mclaughlin04a,chatterjee07a,possenti08a,pellizzoni08a},
but there has been considerable debate over where and how the x--rays are generated.
Normally, pulsar x--ray emission is thought to have a magnetospheric or thermal origin 
\cite{becker00a}.  However, the small binary separation
and interaction of the pulsar wind of the A pulsar with the magnetosphere of the B pulsar
\cite{lyutikov04a} provides alternate x--ray generation mechanisms.

The spin--down luminosity of pulsar A is over three orders of magnitude greater than 
that of pulsar B.  Thus, 
pulsar A is likely to dominate any magnetospheric x--ray emission from the system.
Whilst a wholly magnetospheric (power--law) origin for the observed x--ray emission is 
plausible, the neutral hydrogen column density [$N_{\mathrm H} \sim 1.5\times10^{21}$ 
cm$^{-2}$; \cite{possenti08a}] implied by the model 
is higher than expected from the pulsar's previously assumed location 
in the Gum nebula.  The Gum nebula is believed to be approximately 500 pc distant, with a depth 
of several hundred pc \cite{cordes02a}.

However, the value of 
$N_{\mathrm H}$\ is consistent with the usual average of 10 neutral hydrogen atoms for every
free electron along the line of sight.  
Our revised distance estimate places \dpsr\ beyond the Gum nebula, implying that
the measured value for $N_{\mathrm H}$\ is not discrepant.
It also increases the estimated x--ray luminosity 
by a factor of 5, but the revised value for a power--law fit ($1.2\times10^{31}\,$erg s$^{-1}$)
remains consistent with known relations between pulsar spin--down
luminosity and x--ray luminosity \cite{grindlay02a}.  Hence, our result 
supports a power--law model of
magnetospheric origin (from pulsar A) for the bulk of the x--ray emission from \dpsr.

The discovery of PSR J0737--3039A led to a marked upward revision in the estimated 
Galactic merger rate of DNS systems \cite{kalogera04a}, although uncertainty over the 
characteristics of recycled pulsars means that the true value of the merger rate remains
poorly constrained. Specifically, the distribution of recycled pulsar luminosities
is generally extrapolated from the entire
pulsar population \cite{kalogera04a} even though it (along with the distributions
of pulse shape and beaming fraction) appears to differ from the distribution
for slower pulsars \cite{kramer98a}. 
Our revised distance shows that the radio luminosity of PSR J0737--3039A is a factor of five greater 
than previously assumed.  If this revision were to markedly influence the 
recycled pulsar luminosity function, then the assumed space density of DNS systems 
would be reduced, with a corresponding impact on DNS merger rates estimations.

Finally, we used the measured transverse velocity for \dpsr\ ($24^{+9}_{-6}$\,\kms) 
to constrain models of the formation of the system. 
After subtracting estimates of the peculiar motion of the Solar System 
and Galactic rotation \cite{mignard00a}, we measure a transverse velocity in
local standard of rest of $9^{+6}_{-3}$\,\kms.  
This is comparable to the unadjusted value of 10 \kms\ presented in \cite{kramer06a}, 
and is within the range of transverse velocities expected for the massive stars which
are DNS progenitors [$\sim$20\,\kms; \cite{feast65a}].  
Because the transverse velocity of \dpsr\ is so low, if the
system received a large velocity kick at birth, it must have a large radial
velocity. However, there are no
observational methods available to determine the radial velocity in a DNS system.

Because of the accurate 
measurement of its Shapiro delay, \dpsr\ is known to lie edge--on 
\cite{kramer06a}.  If the only kick it received was
provided by the loss of binding energy during the supernova explosion,
the resultant 3D space velocity should be on the order of $\sim$50 km s$^{-1}$,
estimated from the system's observed eccentricity and orbital velocity  
\cite{blaauw61a}.
This space velocity would be constrained to the plane of the orbit.  From simple
geometry, the probability of observing 
a transverse velocity less than 10 \kms\ is about one in
eight, which is small, but not unreasonable.
Conversely, if the double pulsar had received a large kick \cite{willems06a}, 
the odds of observing such a low transverse velocity become
increasingly remote. Not only would the radial velocity
have to be increasingly large, but the inclination angle of the system must
not be altered by the kick.
Hence, our transverse velocity results reinforce those of
\cite{kramer06a} and are consistent with the interpretation of \cite{piran05a},
who argue for almost no mass loss and kick in the case of \dpsr.

The implication of low kick velocities in \dpsr--like
systems offers a possible, albeit speculative, explanation to the formation of 
PSR J1903+0327, a heavy, highly recycled
millisecond pulsar (mass 1.8 solar masses, period 2.15 ms) 
with a main sequence companion of one solar mass \cite{champion08a}.  
The orbit of such a pulsar should have been highly circularized during the mass--transfer
phase \cite{alpar82a}.  However, PSR J1903+0327 possesses an intermediate orbital 
eccentricity ($e=0.44$).

A formation mechanism for PSR J1903+0327 has been suggested in which a triple system
experiences a white dwarf-neutron star coalescence \cite{van-den-heuvel08a}.
However, a coalescing DNS system such as \dpsr\ could also create a 
PSR J1903+0327--like pulsar.  Thus, given the low velocity of \dpsr, an alternative formation
mechanism for PSR J1903+0327 involves a triple system containing a close DNS binary
and a main sequence star.

\bibliographystyle{science}
\bibliography{deller_thesis}

\begin{thebibliography}{10}
\bibitem{scienceonline}
Information on materials and methods is available on science online.
\end{thebibliography}
29. $\!$The authors acknowledge discussions with M. Kramer and E. van den Heuvel. ATD is supported via a Swinburne University of Technology Chancellor's Research Scholarship and a CSIRO postgraduate scholarship. The Long Baseline Array is part of the Australia Telescope which is funded by the Commonwealth of Australia for operation as a National Facility managed by CSIRO.

\clearpage

\begin{deluxetable}{lc}
\tabletypesize{\tiny}
\tablecaption{Fitted VLBI results for PSR J0737--3039A/B}
\tablewidth{0pt}
\tablehead{ \colhead{Parameter} & \colhead{Value} }
\startdata
Right Ascension $\alpha$ (RA; J2000)\tablenotemark{1}			& 07:37:51.248419(26) \\
Declination $\delta$ (Dec; J2000)\tablenotemark{1}				& $-$30:39:40.71431(10) \\
Proper motion in RA	($\mu_{\alpha}\times\cos{\delta}$; mas yr$^{-1}$)		& $-$3.82(62) \\
Proper motion in Declination ($\mu_{\delta}$; mas yr$^{-1}$) 	& 2.13(23) \\
Parallax (mas)						& 0.87(14) \\
Distance (pc)						& $1150^{+220}_{-160}$ \\
Transverse velocity (km s$^{-1}$)		& $24^{+9}_{-6}$ \\
Transverse velocity in LSR\tablenotemark{2}\ \ (km s$^{-1}$)	& $9^{+6}_{-3}$ \\
Reference epoch (MJD) 				& 54100 \\
\enddata
\tablenotetext{1}{As discussed in \cite{scienceonline}, the actual error in the pulsar position is dominated 
by the alignment of the barycentric reference frame used for pulsar timing and the quasi--inertial 
frame used for VLBI, and is approximately an order of magnitude greater than the formal fit error shown 
here}
\tablenotetext{2}{Local Standard of Rest}
\label{tab:vlbiresults}
\end{deluxetable}

\clearpage

{\bf{\Large{Supporting Online Material}}}

\vspace{4ex}
{\bf{\large{Materials and Methods}}}

\vspace{2ex}
Details of the observational, correlation, and post--processing strategies used are
covered in detail in {\it (S1)} and are summarized here.
Seven observational epochs spanned an 18 month period from August 2006 to February 2008.
Approximately six hours of on--source time (divided equally between the phase reference
calibrator B0736--303 and \dpsr) 
were available per 
epoch, and all available antennas from the Australian Long Baseline Array (LBA) were utilized.  
The LBA is a network of six radio telescopes which provides the best means for VLBI 
study of southern radio sources.  Of these, five telescopes are equipped to make observations
at 1650 MHz, which was the observing frequency utilized.
These were the three ATNF antennas (ATCA, Parkes,
Mopra), the University of Tasmania antenna near Hobart and, when available,
the 70m NASA Deep Space Network (DSN) antenna at the Tidbinbilla station. 
In each session, the on--source time was spread over the ten hour period 
that \dpsr\ was visible to the array in order to obtain good $uv$\ coverage.  Four 16 MHz 
bands were used, quantized at two bit precision to give a total data rate of 256 Mbps.

The phase reference calibrator B0736--303 is a bright and predominantly 
compact radio source separated by $\sim$20 arcminutes from \dpsr.  
Since the position of B0736--303 was not known precisely
in the International Celestial Reference Frame (ICRF) at the outset of this project, 
its position was determined by aligning
the obtained position for \dpsr\ with the value known from timing {\it (S2)}, accounting for the 
pulsar proper motion.  This is possible due to the alignment at the milliarcsecond level of the ICRF
with the barycentric reference frame used for pulsar timing.  The final positional accuracy for
B0736--303 was estimated to be better than 5 milliarcseconds, which is acceptable for the short 
calibrator throw (20 arcminutes) used here. This use of the timing position of \dpsr\ to obtain the 
calibrator position means that the VLBI observations do not contribute to an improvement in the
reference position for \dpsr, as the error in the ultimately obtained position is dominated by the 
uncertainty in alignment between the ICRF and barycentric reference frames 
[1-2 milliarcseconds; {\it(S3)}].  A six minute target/calibrator cycle 
was employed, with observing time divided equally between the target 
and calibrator.

The data were correlated, using matched filtering (gating) on pulse profiles, with the DiFX 
software correlator {\it (S4)}.  Pulse profile filtering improves the signal to noise when observing
a periodic signal by weighting the correlator output according to the varying pulsar flux.  For
\dpsr, the pulse filter was set based on the signal from pulsar A only, since this dominates 
the emission in the system.
The pulse filtering yields a factor of 2.5 improvement in the observed signal to noise ratio.
The most accurate published pulsar ephemeris {\it (S2)} was used to set the 
pulse filter. The visibility data for each epoch was weighted by baseline sensitivity and imaged 
using natural weighting.

The data were fitted using an iterative approach
designed to estimate and account for systematic errors, described in detail in {\it (S1)}.  
This approach generated a position and position error for each 16 MHz band at each epoch, and
the single--band positions were used to obtain a single weighted centroid position for each 
epoch.  Estimates of systematic error at each epoch were made by inspecting the reduced 
chi--squared of the fit to position centroid, and these estimates were added in quadrature to the
centroid position error. The final seven position measurements were then fit for position, proper 
motion and parallax using the iterative minimization code described in {\it (S1)}. The reduced
chi--squared of the final fit was 0.79, indicating that the quoted errors for the fitted values 
may be slightly conservative.

\clearpage

\vspace{4ex}
{\bf{\large{SOM References and Notes}}}

\vspace{2ex}
S1. A. T. Deller, S. J. Tingay, W. F. Brisken, {\it ApJ} {\bf 690}, 198 (2009).

S2. M. Kramer, {\it et al.}, {\it Science} {\bf 314}, 97 (2006).

S3. A. T. Deller, J. P. W. Verbiest, S. J. Tingay, M. Bailes, {\it ApJL} {\bf 685}, L67 (2008).

S4. A. T. Deller, S. J. Tingay, M. Bailes, C. West, {\it PASP} {\bf 119}, 318 (2007).

\end{document}